\documentclass[12pt]{iopart}

\usepackage{harvard}
\usepackage{amssymb}
\usepackage{amsthm}

 \newtheorem{theorem}{Theorem}[section]
       
       \newtheorem{proposition}[theorem]{Proposition}
       
    \newtheorem{definition}[theorem]{Definition}
       
\begin{document}

\title[]{Maximum Entropy in the framework of Algebraic Statistics: A
  First Step}
\author{Ambedkar Dukkipati}
\address{EURANDOM, P.O. Box 513, 5600 MB Eindhoven, The Netherlands}
\ead{\mailto{dukkipati@eurandom.tue.nl}
}

\begin{abstract}
Algebraic statistics is a recently evolving field, where one would treat
statistical models as algebraic objects and thereby use tools from
computational commutative algebra and algebraic geometry in the
analysis and computation of statistical models. 
In this approach, calculation of 
parameters of statistical models amounts to solving set of
polynomial equations in several variables, for which one can use
celebrated Gr\"{o}bner bases theory.  
Owing to the important role of information theory 
in statistics, this paper as a first step, explores the possibility
of describing  maximum and minimum entropy (ME) models in the
framework of algebraic statistics.
We show that ME-models are toric models (a class of
algebraic statistical models) when the constraint functions (that
provide the information about the underlying random variable) are
integer valued functions, and the set of statistical models that
results from ME-methods are indeed an affine variety. 

\end{abstract}

\maketitle

\section{Introduction}
\label{Section:Introduction}
       \noindent
       Algebra has always played an important role in
       statistics, a classical example being linear algebra.
      There are also many other instances of applying algebraic
       tools in
              statistics~(e.g~\cite{VianaRichards:2001:AlgebraicMethodsInStatisticsAndProbability}). But,
       treating statistical models as algebraic objects,
       and thereby using tools of computational
       commutative algebra and algebraic geometry in the analysis of
       statistical models is very recent and has led to the still evolving
       field of {\em algebraic statistics}. 

       The use of computational algebra and algebraic geometry in
       statistics was initiated in
       the work of Diaconis and
       Sturmfels~\cite{DiaconisSturmfels:1998:AlgebraicAlgorithmsForSamplingFromConditionalDistributions}
       on exact hypothesis tests of conditional independence in
       contingency tables,
       and in the work of Pistone et
       al.~\cite{PistoneRiccomagnoWynn:2001:AlgebraicStatistics} in
       experimental design. The term `Algebraic Statistics' was first
       coined in the
       monograph by Pistone et al.~\cite{PistoneRiccomagnoWynn:2001:AlgebraicStatistics}
       and appeared recently in the title of the
       book by Pachter and
       Sturmfels~\cite{PachterSturmfels:2005:AlgebraicStatisticsAndComputationalBiology}.  

       To extract the underlying algebraic
       structures in discrete statistical models,  algebraic
       statistics treat 
       statistical models  as affine varieties. (An affine variety is
       the set of all solutions to family of polynomial equations.) 
       Parametric statistical models are described in terms of a
       polynomial (or rational) mapping from a set of parameters to
       distributions. One can show that many statistical models, for
       example independence models, Bernouli random variable etc. (see
       ~\cite{PachterSturmfels:2005:AlgebraicStatisticsAndComputationalBiology}
       for more examples), can be given this algebraic formulation, and these are
       referred to as algebraic statistical models.

       Exponential models, which form the important class of statistical
       models, are
       studied in
       algebraic statistics under the name `toric' models by using
       maximum likelihood methods.
       Toric
       models are algebraic statistical models and the term `toric'
       comes from important algebraic objects known as `toric ideals' in
       computational algebra. 
       In this view of very established
       role of information theory in
       statistics~\cite{Kullback:1959:InformationTheoryAndStatistics,CsiszarShields:2004:InformationTheoryAndStatistics}
       this paper attempts to describe maximum entropy models in
       algebraic statistical framework.
  
       In particular, we show that maximum entropy models (also minimum
       relative-entropy models) are indeed toric models, when the
       functions that provide the information about the underlying
       random variable in the form of expected values are integer
       valued. We also show that when the information is available in
       the form of sample means, by modifying maximum entropy
       prescriptions  calculating model parameters amounts to solving
       set of polynomial equations. This establishes a fact that
       set of statistical models results from maximum entropy
       methods are indeed algebraic varieties.

       A note on the results presented in this paper: we will not
       present the details on 
       Gr\"{o}bner bases theory and related concepts to solve the polynomial equations due
       to space constraint; we refer reader to text books on
       computational algebra and Gr{\"{o}}bner basis
       theory~\cite{AdamsLoustaunau:1994:AnIntroductionToGrobnerBases,CoxLittleOshea:1997:IdealsVarietiesAndAlgorithms}.



       We organize our paper as follows.
       In \S~\ref{Section:AlgebraicStatisticalModels} we give basic
       notions of algebra and introduce notation along with an
       introduction to algebraic statistics.
       \S~\ref{Section:MEinAlgebraicStatisticalSetup} describes
       maximum entropy (ME) prescriptions in algebraic statistical
       framework by introducing important algebraic objects called
       toric ideals. In
       \S~\ref{Section:calculationOfMEdistributionsViaPolynomialEquations}
       we show how one can transform the problem of calculating ME
       distributions to solving set of polynomial equations.
       
\section{Algebraic Statistical Models}
\label{Section:AlgebraicStatisticalModels}
   \subsection{Basic notions of Algebra}
       \noindent
       Through out this paper $k$ represents a field. A monomial in
       $n$ indeterminates $x_{1}, \ldots, x_{n}$ is a power product of the
       form $x_{1}^{\alpha_{1}} \ldots x_{n}^{\alpha_{n}}$, where all
       the exponents are nonnegative integers, i.e.  
       $\alpha_{i} \in {\mathbb{Z}}_{\geq 0}$, $i = 1, \ldots n$. One
       can simplify the notation for monomial as follows: denote $\alpha
       = (\alpha_{1}, \ldots, \alpha_{n}) \in {\mathbb{Z}}_{\geq
         0}^{n}$ and by using multi-index notation we set
       \begin{displaymath}
         x^{\alpha} = x_{1}^{\alpha_{1}} \ldots x_{n}^{\alpha_{n}}
       \end{displaymath}
       with the understanding that $x=(x_{1}, \ldots, x_{n})$. Note
         that $x^{\alpha} =1$ when ever 
         $\alpha=(0,\ldots,0)$. 
       Once the order of the indeterminates are
       fixed, monomial  $x_{1}^{\alpha_{1}} \ldots x_{n}^{\alpha_{n}} =
       x^{\alpha}$ is identified by $(\alpha_{1}, \ldots,
       \alpha_{n})$. Hence, set of all monomials in indeterminates
       $x_{1}, \ldots, x_{n}$ can be represented by $\mathbb{Z}_{\geq
         0}^{n}$. Theory of monomials is central to the celebrated Gr{\"{o}}bner
       bases theory in computational algebra which provides tools for  
       solving set of polynomial equations and related
       problems in algebraic
       geometry~\cite{MishraYap:1989:NotesOnGrobnerBases}.   
       Monomial theory  itself plays important role in 
       algebraic statistics in the representation of exponential models
       where probabilities are expressed in terms of power
       products~\cite{Rapallo:2006:ToricStatisticalModels}.

       A polynomial $f$ in $x_{1}, \ldots, x_{n}$ with coefficients in $k$
       is a finite linear combination of monomials and can be written in
       the form of
       \begin{displaymath}
       f = \sum_{\alpha \in \Lambda_{f}} a_{\alpha} x^{\alpha} \enspace,
       \end{displaymath}
       where $\Lambda_{f} \subset {\mathbb{Z}}_{\geq 0}^{n}$ is a
       finite set and
       $a_{\alpha} \in k$. The collection of all polynomials in the
       indeterminates $x_{1}, \ldots, x_{n}$ is the set $k[x_{1}, \ldots,
       x_{n}]$ and it has structure not only
       of a vector space but also of a ring. Indeed the ring
       structure of $k[x_{1},\ldots,x_{n}]$ plays main role in
       computational algebra and 
       algebraic geometry.

       A subset $\mathfrak{a} \subset k[x_{1},\ldots,x_{n}]$ is said to
       be ideal if it satisfies: (i) $0 \in \mathfrak{a}$ (ii) $f,g \in
       \mathfrak{a}$, then $f+g \in \mathfrak{a}$ (iii) $f \in
       \mathfrak{a}$ and $h \in  k[x_{1},\ldots,x_{n}]$ and then $hf \in
       \mathfrak{a}$. 
       A set $V \subset k^{n}$ is said to be affine variety if there
       exists $f_{1}, \ldots, f_{s} \in k[x_{1}, \ldots, x_{n}]$ such
       that
       \begin{displaymath}
         V = \{(c_{1},\ldots c_{n}) \in k^{n} : f_{i}(c_{1},\ldots
         c_{n}) =0, 1 \leq i \leq s \} \enspace.
       \end{displaymath}
       We use the notation $\mathcal{V}(f_{1}, \ldots, f_{s}) =V$.

   \subsection{Algebraic Statistical Model}
       \noindent
       At the very core of the field of algebraic statistics lies the
       notion of an `algebraic statistical model'. While this notion has
       the potential of serving as a unifying theme for algebraic
       statistics, there is no unified definition of an algebraic
       statistical
       model~\cite{DrtonSullivant:2006:AlgebraicStatisticalModels}. Here,
       we adopt the appropriate definition of statistical
       model
       from~\cite{PachterSturmfels:2005:AlgebraicStatisticsAndComputationalBiology,DrtonSullivant:2006:AlgebraicStatisticalModels}. For
       a recent elaborate discussion on formal definition of algebraic
       statistical models one can refer
       to~\cite{DrtonSullivant:2006:AlgebraicStatisticalModels}.   
   
       Let $X$ be a discrete random variable taking finitely many values from the
       set $[m] = \{1,2,\ldots m \}$. A probability distribution $p$
       of $X$  is naturally represented as a vector $p=(p_{1}, \ldots,
       p_{m}) \in {\mathbb{R}}^{m}$ if we fix the order on $[m]$.
       Then set of all probability mass
       functions (pmfs) of $X$ is called probability simplex
       \begin{equation}
         \Delta_{m-1} = \{ p= (p_{1},\ldots,p_{m}) \in
               {\mathbb{R}}_{\geq 0}^{m} : \sum_{i=1}^{m} p_{i} =1 \} \enspace.
       \end{equation}
       The index $m-1$ indicates the dimension of the simplex $\Delta_{m-1}$.
       A statistical model $\mathcal{M}$ is a subset of
       $\Delta_{m-1}$ and is
       said to be algebraic if $\exists f_{1}, \ldots, f_{s} \in
       k[p_{1},\ldots,p_{m}]$ such that
       \begin{displaymath}
          \mathcal{M} = \mathcal{V} (f_{1}, \ldots, f_{s}) \cap
          \Delta_{m-1} \enspace.
       \end{displaymath}
       Now we move on to parametric statistical models and their
       algebraic formulations. 
       
       Let $\Theta \subseteq \mathbb{R}^{d}$ be a
       parametric space and $\kappa: \Theta \rightarrow \Delta_{m-1}$
       be a map. The image $\kappa(\Theta)$ is called parametric
       statistical model. Given a statistical model $\mathcal{M}
       \subseteq \Delta_{m-1}$, by parametrization of $\mathcal{M}$ we
       mean, identifying a set $\Theta \subseteq \mathbb{R}^{d}$ and a
       function $\kappa
       : \Theta \rightarrow \Delta_{m-1}$ such that $\mathcal{M} =
       \kappa({\Theta})$. To describe more general statistical models
       in algebraic framework we need  following notion of
       {\em semi-algebraic set}.
       
       \begin{definition}
       \label{Definition:SemiAlgebraicSet}  
       A set $\Theta
       \subseteq \mathbb{R}^{d}$ is called semi-algebraic set, if
       there are two finite collection of polynomials $F \subset k[x_{1},\ldots,x_{d}]$
       and $G \subset k[x_{1},\ldots,x_{d}]$ such that
       \begin{displaymath}
         \Theta = \{ \theta \in \mathbb{R}^{d}: f(\theta) =0, \forall
         f \in F \:\mbox{and}\: g(\theta) \geq 0, g \in G \} \enspace.
       \end{displaymath}
       \end{definition}
       Now we have following definition of parametric algebraic
       statistical model.
       \begin{definition}
        \label{Definition:ParametricAlgebraicStatisticalModel} 
         Let $\Delta_{m-1}$ be a probability simplex and $\Theta
         \subset \mathbb{R}^{d}$ be a semi-algebraic set. Let $\kappa:
         \mathbb{R}^{d} \rightarrow \mathbb{R}^{m}$ be a rational
         function (a rational function is a quotient of two polynomials)
         such that $\kappa(\Theta) \subseteq \Delta_{m-1}$. Then the
         image $\mathcal{M} = \kappa(\Theta)$ is a parametric
         algebraic statistical model.
       \end{definition}
       
         Conversely, a parametric
         statistical model $\mathcal{M} = \kappa(\Theta) \subseteq
         \Delta_{n-1}$ is said to be algebraic if $\Theta$ is
         semi-algebraic set and $\kappa$ is a rational
         function. 
       From now on we refer to `parametric algebraic statistical models'
       as `algebraic statistical models'. 

       In this paper we consider following special case 
       of algebraic statistical models (cf.~\cite[pp
         7]{PachterSturmfels:2005:AlgebraicStatisticsAndComputationalBiology}). Consider 
       a map 
       \begin{eqnarray}
       \label{Equation:PolynomialAlgebraicStatisticalModel}  
       \kappa: \Theta (\subseteq {\mathbb{R}}^{d}) \rightarrow {\mathbb{R}}^{m}
       \nonumber \\
        \kappa: \theta=(\theta_{1},\ldots,\theta_{d}) \mapsto
       \left(\kappa_{1}(\theta),\ldots, \kappa_{m}(\theta)  \right) 
       \end{eqnarray}
       where $\kappa_{i} \in k[\theta_{1},\ldots,\theta_{d}]$. We
       assume that $\Theta$ satisfies
       $\kappa_{i}(\theta) \geq 0$, $i=1,\ldots,m$
       and $\sum_{i=1}^{m} \kappa_{i}(\theta) =1$ for any $\theta \in \Theta$.   
       Under these
       conditions $\kappa(\Theta)$ is indeed an algebraic statistical
       model (Definition~\ref{Definition:ParametricAlgebraicStatisticalModel}) since 
       $\kappa(\Theta) \subset \Delta_{m-1}$, $\kappa$ is a polynomial 
       function and $\Theta$ is a semi-algebraic set ($H= \left\{
       \sum_{i=1}^{m} f_{i} - 1 \right\}$ and $G= 
       \left\{ f_{i}: i =1,\ldots, m\right\}$ in the
       Definition~\ref{Definition:SemiAlgebraicSet}).  

       Some statistical models are naturally given by a polynomial map
       $\kappa$ (\ref{Equation:PolynomialAlgebraicStatisticalModel})
       for which the condition $\sum_{i=1}^{m} \kappa_{i}(\theta) =1$
       does 
       not hold. If this is the case one can consider following
       algebraic statistical model:
       \begin{equation}
       \label{Equation:RationalAlgberaicStatisticalModel}  
       \kappa: \theta=(\theta_{1},\ldots,\theta_{d}) \mapsto
       \frac{1}{\sum_{i=1}^{m}\kappa_{i}(\theta)}\left(\kappa_{1}(\theta),\ldots,
       \kappa_{m}(\theta)  \right) \enspace,
       \end{equation}
       assuming that remaining conditions that have been specified for the model
       (\ref{Equation:PolynomialAlgebraicStatisticalModel}) are valid here 
       too. The only difference is that instead of $\kappa$ being a
       polynomial map, we have it as a rational map.


\section{ME in algebraic statistical setup}
\label{Section:MEinAlgebraicStatisticalSetup}
   \subsection{Toric Models}
   \noindent
       In the algebraic description of exponential models monomials
        and binomials play a fundamental role.
       The study of relations of power products
       lead to the theory of toric ideals in the commutative
       algebra~\cite{Sturmfels:1996:GrobnerBasisAndConvexPolytopes}.
       Here we describe basic notion of toric
       ideal that are relevant to representation and computation of
       discrete exponential models; for more details on theory and
       computation of toric 
       ideals one can refer
       to~\cite{Sturmfels:1996:GrobnerBasisAndConvexPolytopes,BigattiRobbianto:2001:ToricIdeals,BigattiLa-ScalaRobbianto:1999:ComputingToricIdeals}.

       Before we give the definition of toric ideal, we describe the
       notion of Laurent polynomial.  
       If we allow negative
       exponents in a polynomial i.e., polynomial of the form
       $f= \sum_{\alpha \in \Lambda_{f}} a_{\alpha} x^{\alpha}$ where $\alpha
       \in {\mathbb{Z}}^{n}$, it is known as Laurent polynomial
       ($\Lambda_{f} \subset {\mathbb{Z}}_{\geq 0}^{n}$ is finite). Set of
       all Laurent polynomials in the indeterminates $x_{1}, \ldots,
       x_{n}$ is denoted by
       $k[x_{1}^{\pm},\ldots,x_{n}^{\pm}]$ and it also has a structure of a
       ring.
       
       Now we define the toric ideal.  
       \begin{definition}
       \label{Definition:ToricIdeal}
         Let $A=[a_{ij}] \in {\mathbb{Z}}^{d \times n}$ be a matrix
         with rank $d$.
         Consider the ring homeomorphism
         \begin{eqnarray}
          \label{Equation:ToricMap} 
           \hat{\pi}: k[x_{1},\ldots,x_{n}] \rightarrow
           k[\theta_{1}^{\pm},\ldots,\theta_{d}^{\pm}] \nonumber
           \\
           \hat{\pi}: x_{j} \mapsto \theta_{1}^{a_{1j}} \ldots\theta_{1}^{a_{dj}}
          \end{eqnarray}
         The toric ideal ${\mathfrak{a}}_{A}$  of $A$ is defined as 
         the kernel of the map $\hat{\pi}$, i.e., ${\mathfrak{a}}_{A}
         = \mbox{ker} \hat{\pi}$.
       \end{definition}
       
       The mapping $\hat{\pi}$ can be viewed as ``parametrization'' and
       which can be explained by the following description of
       $\hat{\pi}$. Consider a map
       \begin{eqnarray}
        \label{Equation:MapThatLiftsToToricMap} 
       \pi: \mathbb{Z}_{\geq 0}^{n} \rightarrow \mathbb{Z}^{d} \nonumber \\
       \pi: u=(u_{1}, \ldots, u_{n}) \mapsto Au .
       \end{eqnarray}
       The map $\pi$ lifts to the ring homomorphism $\hat{\pi}$ in the
       sense of action of $\hat{\pi}$ on
       $x^{u} = x_{1}^{u_{1}} \ldots x_{n}^{u_{n}}
       \in k[x_{1}, \ldots, x_{n}]$. That is
      \begin{eqnarray}
        \hat{\pi} (x^{u}) = \hat{\pi} (x_{1}^{u_{1}}, \ldots,
        x_{n}^{u_{n}}) 
        = {\left(\prod_{i=1}^{d} \theta_{i}^{a_{i1}} \right)}^{u_{1}}
        \ldots {\left( \prod_{i=1}^{d} \theta_{i}^{a_{in}} 
            \right)}^{u_{n}} \\
        = \prod_{i=1}^{d} \theta_{i}^{\sum_{j=1}^{n} a_{ij}u_{j}}
        = \theta^{Au}  \enspace. 
       \end{eqnarray}
       
 
       Toric ideal theory plays an important role in applications of
       computational algebraic geometry like integer
       programming
       etc.(cf.~\cite{Sturmfels:1996:GrobnerBasisAndConvexPolytopes}). Note
       that in the algebraic descriptions of exponential models and
       their maximum likelihood estimates only non-negative cases of
       toric ideals (and hence toric models) is considered i.e., the
       matrix $A=[a_{ij}]$ in Definition~\ref{Definition:ToricIdeal}
       is assumed to be nonnegative and 
       the map (\ref{Equation:ToricMap}) is specified as $\hat{\pi}:
       k[x_{1},\ldots,x_{n}] \rightarrow k[\theta_{1},\ldots,\theta_{d}]$
           (see~\cite{PachterSturmfels:2005:AlgebraicStatisticsAndComputationalBiology}). As
       described later in this paper, in the algebraic descriptions 
       of maximum entropy models one has to deal with the Laurent
       polynomials and hence one has to include the negative case in
       the definitions of toric ideals and toric models. This poses
       no problem because toric ideal theory in commutative algebra
       naturally includes the negative case (as in
       Definition~\ref{Definition:ToricIdeal}) and Gr\"{o}bner bases
       theory can be extended to Laurent polynomial
       ring~\cite{PauerUnterkircher:1999:GrobnerBasesForIdealsInLaurentPolynomialRings}.
       
       The concept of toric ideals let to the description of
       exponential models under the name toric models in algebraic
       statistics which is defined as follows.
       \begin{definition}
       \label{Definition:ToricModel} 
         Let $A \in \mathbb{Z}_{\geq 0}^{d \times m}$ be a matrix such that the
         vector $(1, \ldots, 1) \in \mathbb{Z}_{\geq 0}^{m}$ is in the
         row span of $A$. Let $h \in \mathbb{R}_{> 0}^{m}$ be a vector
         of positive real numbers. Let $\Theta = \mathbb{R}_{> 0}^{m}$
         and let $\kappa^{A,h}$ be the rational parametrization
         \begin{eqnarray}
         \label{Equation:ToricModels} 
         \kappa^{A,h}: \Theta &\rightarrow& {\mathbb{R}}^{m} \nonumber
         \\
         \kappa_{j}^{A,h}: \theta &\mapsto& Z(\theta)^{-1}
         h_{j}  \prod_{i=1}^{d} \theta_{i}^{a_{ij}} \enspace,
         \end{eqnarray}
         where $\theta = (\theta_{1}, \ldots, \theta_{d})$ and
         $Z(\theta)$ is the appropriate normalizing constant.
         The toric model is the parametric algebraic statistical model
         \begin{equation}
           {\mathcal{M}}_{A,h} \triangleq \kappa^{A,h}(\Theta) \enspace.
         \end{equation}  
       \end{definition}
       Independence models, exponential models, Markov chains and
       Hidden Markov chains can be given an algebraic statistical
       description by means of toric
       models~\cite{PachterSturmfels:2005:AlgebraicStatisticsAndComputationalBiology}. We
       keep positivity of $A$ in the
       Definition~\ref{Definition:ToricModel} as a matter of
       convention.
       
  \subsection{ME in terms of Toric Models}
  \noindent
       Let $X$ be a random variable taking values from the set $[m] =
       \{1,2,\ldots m \}$. The only information we know about the
       pmf $p = (p_{1}, \ldots, p_{m})$ of $X$ is in the form of expected
       values of the functions $t_{i}:[m] \rightarrow \mathbb{R}$, $i=1,\ldots,
       d$ (we refer these functions as `constraint functions'). We
       therefore have  
       \begin{equation}
        \label{Equation:ExpectationConstrainsForIntegerValuedFunctions} 
         \sum_{j=1}^{m} t_{i}(j) p_{j}= T_{i}  \enspace, i = 1, \ldots
         d \enspace,
       \end{equation}
       where $T_{i}$, $i=1,\ldots,d$, are assumed to be known. In
       an information theoretic approach to statistics, known as
       Jayens maximum entropy model, one would
       choose the pmf $p \in \Delta_{m-1}$ that maximize the 
       Shannon entropy functional
       \begin{equation}
         S(p) = - \sum_{j=1}^{m} p_{j} \ln p_{j}
       \end{equation}  
       with respect to the constraints
       (\ref{Equation:ExpectationConstrainsForIntegerValuedFunctions}).
              
       The corresponding Lagrangian can be written
       as
       \begin{equation}
       \label{Equation:LagrangianForMaximumEntropy}	 
         \Xi(p,\xi) \equiv S(p) - \xi_{0}\left( \sum_{j=1}^{m}p_{j} -1
         \right) -  \sum_{i=1}^{d} \xi_{d}
         \left( \sum_{j=1}^{m} t_{i}(j) p_{j} - T_{i}  \right)
       \end{equation}  
       Holding $\xi=(\xi_{1},\ldots,\xi_{d})$ fixed, the unconstrained maximum of Lagrangian
       $\Xi(p,\xi)$ over all $p \in \Delta_{m-1}$ is given by an
       exponential
       family~\cite{CoverThomas:1991:ElementsOfInformationTheory}
       \begin{equation}
       \label{Equation:MaximumEntropyDistribution}
        p_{j}(\xi) = Z(\xi)^{-1}\exp{\left( - \sum_{i=1}^{d} \xi_{i} t_{i}(j)
          \right)} \enspace, j = 1, \ldots, m, 
       \end{equation}
       where $Z(\xi)$ is normalizing constant given by
       \begin{equation}
         Z(\xi) = \sum_{j=1}^{m} \exp{\left( - \sum_{i=1}^{d} \xi_{i} t_{i}(j)
          \right)} \enspace.
       \end{equation} 
       For various values of $\xi \in \mathbb{R}^{d}$, the family
       (\ref{Equation:MaximumEntropyDistribution}) is known as {\em maximum
       entropy model}.
       

       Now, we have following proposition.
       \begin{proposition}
       \label{Propostion:ME-ModelAsToricModel} 	 
         The maximum entropy model
         (\ref{Equation:MaximumEntropyDistribution}) is a toric model
         provided that the constraint functions 
         are integer valued. 
       \end{proposition}  
       \proof
       Set $\ln
        \theta_{i} = \xi_{i}$, $i=1,\ldots,d$. Now,
        (\ref{Equation:MaximumEntropyDistribution}) gives us
        \begin{eqnarray}
	\label{Equation:ParametrizationOfME-model}  
          p_{j} = Z(\theta)^{-1} \exp{\left( - \sum_{i=1}^{d}  t_{i}(j) \ln \theta_{i}
          \right)} 
                = Z(\theta)^{-1} \prod_{i=1}^{d} \theta_{i}^{t_{i}(j)} \enspace.
        \end{eqnarray}	
        By defining matrix $A = [t_{i}(j)] \in {\mathbb{Z}}^{d \times
        n}$ and setting $h = (\frac{1}{m},\ldots,\frac{1}{m})$ we have
        rational parametrization as in (\ref{Equation:ToricModels}). 
       \endproof 
        Note that we allowed only integer valued functions in the 
        ME-model in the above proposition, which is
       necessary for algebraic descriptions of the same.
       Here we also mention that in the above proof by assuming $h \in
       \Delta_{m-1}$ (which acts as a prior), we can imply that
       minimum I-divergence model~\cite{Csiszar:1975:I-devergenceOfProbabilityDistributionsAndMinimizationProblems}
       \begin{equation}
       \label{Equation:MinimumEntropyDistribution}
        p_{j} = {\widehat{Z}(\xi)}^{-1} h_{j} \exp{\left( - \sum_{i=1}^{d} \xi_{i} T_{i}(j)
          \right)} \enspace,\:\: j = 1, \ldots, m, 
       \end{equation}
       (with appropriate normalizing constant $\widehat{Z}(\xi)$)
       is indeed a toric model.

       Once the specification of statistical model is done, the task
       is to calculate the model parameters with the available
       information. In this case the available information is in the form of expected
       valued of functions $t_{i}$, $i=1,\ldots d$ and
       the Lagrange parameters $\xi_{i}$, $i=1,\ldots, d$ are
       determined using the constrains
       (\ref{Equation:ExpectationConstrainsForIntegerValuedFunctions}). 

\section{Calculation of ME distributions via solving Polynomial
  equations}
\label{Section:calculationOfMEdistributionsViaPolynomialEquations}
   \subsection{Direct method}
      \noindent
      One can show that the Lagrange parameters in ME-model
       (\ref{Equation:MaximumEntropyDistribution}) can be estimated
       by solving following set of
       partial differential equations~\cite{Jaynes:1968:PriorProbabilities}
       \begin{equation}
         \frac{\partial}{\partial \xi_{i}} \ln Z(\xi) = 
          T_{i} \enspace, i=1,\ldots, d ,
       \end{equation}
       which has no explicit analytical solution.
       In literature there are several methods of estimating
       ME-models. One of the important method is Darroch and Ratcliff's
       generalized iterative scaling
       algorithm~\cite{DarrochRatcliff:1972:GeneralizedIterativeScalingForLogLinearModels}. 
       Here we can show that ME-models can be calculated using
       computational algebraic methods. 
   
       Note that set of all distributions which satisfies
       (\ref{Equation:ExpectationConstrainsForIntegerValuedFunctions})
       is known as linear family (we denote this by
       $\mathcal{L}$). Now, if we represent the exponential
       family (\ref{Equation:MaximumEntropyDistribution}) by
       $\mathcal{E}$, the set of statistical models that results from
       ME-principle can be written as $\mathcal{L} \cap \mathcal{E}
       \subset \Delta_{m-1}$. One can show that $\mathcal{L} \cap \mathcal{E}
       \subset \Delta_{m-1}$ is a variety.

       By substituting maximum entropy
       distributions (\ref{Equation:ParametrizationOfME-model}) in
       (\ref{Equation:ExpectationConstrainsForIntegerValuedFunctions})
       we get
       \begin{equation}
	 \sum_{j=1}^{m} t_{i}(j) \prod_{i=1}^{d} \theta_{i}^{t_{i}(j)}
	 = T_{i} Z(\theta) \enspace,
       \end{equation}
       which can be written as
       \begin{equation}
       \label{Equation:MaximumEntropyPolynomialEquation}	 
	\sum_{j=1}^{m} t_{i}(j) \prod_{i=1}^{d} \theta_{i}^{t_{i}(j)}
	 = T_{i} \sum_{j=1}^{m}  \prod_{i=1}^{d} \theta_{i}^{t_{i}(j)}  \enspace.
       \end{equation}
       The solutions of system of polynomial equations
       (\ref{Equation:MaximumEntropyPolynomialEquation}) gives the
       maximum entropy model spcified the available information
       (\ref{Equation:ExpectationConstrainsForIntegerValuedFunctions}).
       We state this as a proposition.

       \begin{proposition}
	  The maximum entropy model
         (\ref{Equation:MaximumEntropyDistribution}) can be specified
         by solving set of polynomial equations provided that the
         constraint functions $t_{i}$, $i=1,\ldots, d$ 
         are integer valued.
       \end{proposition}
 
  \subsection{Dual Method}
  \noindent
         Here we follow the method of {\em dual} optimization problem.
       By using Kuhn-Tucker theorem we calculate Lagrange parameters
       $\xi_{i}$, $i=1,\ldots,d$
       in~(\ref{Equation:MaximumEntropyDistribution}) by optimizing
       dual  of 
       $\Xi(p,\xi)$. That is the task is to find $\xi$ which maximizes
       \begin{equation}
       {\Psi}(\xi) \equiv \Xi(p^{(\xi)},\xi) \enspace.
       \end{equation}
       Note that $\Psi(\xi)$ is nothing but entropy of
       ME-distribution~(\ref{Equation:MaximumEntropyDistribution}). We have
       \begin{equation}
         \Psi(\xi) = \ln Z + \sum_{i=1}^{d} \xi_{i} T_{i} \enspace.
       \end{equation}
       This can be written as
         \begin{eqnarray}
         \Psi(\xi) &=& \ln \sum_{j=1}^{m} \exp \left(- \sum_{j=1}^{d}
         \xi_{i}t_{i}(j)  \right)    + \sum_{i=1}^{d} \xi_{i} T_{i}
         \nonumber \\
         &=& \ln  \sum_{j=1}^{m} \exp \left( \xi_{i} (T_{i} -
         t_{i}(j))  \right) \enspace.
         \end{eqnarray}
         Now maximizing $\Psi(\xi)$ is equivalent to maximizing
         \begin{equation}
           \Psi'(\xi)  = \sum_{j=1}^{m} \exp \left( \xi_{i} (T_{i} -
           t_{i}(j))  \right) \enspace.
         \end{equation}
         By introducing $\xi_{i} = \ln \theta_{i}$, $i=1,\ldots,d$ we have
         \begin{equation}
         \Psi'(\theta) = \sum_{j=1}^{m} \prod_{i=1}^{d}
         \theta_{i}^{T_{i} - t_{i}(j)} \enspace.
         \end{equation}
         The solution is given by solving the following set of
         equations
         \begin{equation}
           \frac{\partial \Psi'}{\partial \theta_{j}} =0 \enspace,
           j=1,\ldots d .
         \end{equation}
         Unfortunately $\frac{\partial \Psi}{\partial \theta_{j}} \in
         k[\theta_{1}^{\pm},\ldots, \theta_{d}^{\pm}] $ only if $T_{i}
         \in \mathbb{Z}$. Now, we consider the case where the expected
         values are available as sample means.
         
      In most practical problems the information in the form of
      expected values
      is available via sample or empirical means. That is, given a
      sequence of observations $O_{1}, \ldots, O_{N}$ the sample
      means ${\widetilde{T}}_{i}$, $i=1,\ldots,d$, with respect to the
      functions $t_{i}$, $i=1,\ldots, d$ are given by
       \begin{equation}
       \label{Equation:SampleMean} 
        {\widetilde{T}}_{i} = \frac{1}{N} \sum_{l=1}^{N}
      t_{i}(O_{l}), i=1, \ldots, d,
      \end{equation}
      and the underlying hypothesis is $T_{i} \approx {\widetilde{T}}_{i}$. That is
      \begin{equation}
       \label{Equation:SampleMeanHypothesis}  
        \sum_{j=1}^{m}p_{j}t_{i}(j)  \approx \frac{1}{N}
        \sum_{l=1}^{N} t_{i}(O_{l}) \enspace, i=1, \ldots, d.
      \end{equation}  
      Now we show that, by choosing alternate Lagrangian in the place
      of (\ref{Equation:LagrangianForMaximumEntropy}) we can
      transform the parameter estimation of ME-model to a problem of solving
      set of polynomial (Laurent) equations. 
      \begin{proposition}
       Given the hypothesis (\ref{Equation:SampleMeanHypothesis}) the
       problem of 
       estimating the ME-model in the dual method
       amounts to solving set of Laurent polynomial equations
       (assuming that constraint functions are integer valued). 
      \end{proposition}
      \proof
      To retain the integer valued exponents in our final solution we
      consider the constrains of  the form
       \begin{equation}
        \label{Equation:ModifiedExpectationConstrainsForIntegerValuedFunctions} 
         N\sum_{j=1}^{m} t_{i}(j) p_{j}= \sigma_{i}  \enspace, \:\:\:i = 1, \ldots
         d \enspace,
       \end{equation}
       where $\sigma_{i} = \sum_{l=1}^{N } t_{i}(O_{l}) $ denotes the sample sum.       
      In this case Lagrangian is
      \begin{equation}
         \widetilde{\Xi}(p,\xi) \equiv S(p) - \xi_{0}\left( \sum_{j=1}^{m}p_{j} -1
         \right)  - \sum_{i=1}^{d} \widetilde{\xi}_{d}
          \left(N \sum_{j=1}^{m} p_{j} t_{i}(j)  - \sigma_{i}
         \right) \enspace.
      \end{equation}  
      This results in the ME-distribution
      \begin{equation}
       \label{Equation:MaximumEntropyDistribution_BasedOnEmpiricalMeans}
        p_{j}(\xi) = \widetilde{Z}(\xi)^{-1}\exp{\left( - N \sum_{i=1}^{d} \widetilde{\xi}_{i} t_{i}(j)
          \right)} \enspace, \:\:\: j = 1, \ldots, m, 
      \end{equation}
       where $Z(\xi)$ is normalizing constant given by
       \begin{equation}
         \widetilde{Z}(\xi) = \sum_{j=1}^{m} \exp{\left( - N \sum_{i=1}^{d} \widetilde{\xi}_{i} t_{i}(j)
          \right)} \enspace.
       \end{equation}
       To calculate the parameters we maximize the dual $\widetilde{\Psi}(\widetilde{\xi})$
       of $\widetilde{\Xi}(p,\xi)$.
       That is we maximize the functional
       \begin{eqnarray}
         \widetilde{\Psi}(\widetilde{\xi}) = \ln \widetilde{Z} + \sum_{i=1}^{d}
         {\widetilde{\xi}}_{i} {\sigma}_{i} \enspace.
        \end{eqnarray}
       It is equivalent to optimizing the functional
       \begin{displaymath}
        \widetilde{\Psi}'(\widetilde{\xi}) = \sum_{j=1}^{m} \exp \left( \sum_{i=1}^{d} \widetilde{\xi_{i}}
         \sigma_{i} - N \sum_{i=1}^{d} \widetilde{\xi_{i}} t_{i}(j) \right)
       \end{displaymath}
       By setting $\ln \widetilde{\theta}_{i} = {\widetilde{\xi}}_{i}$ we have
        \begin{equation}
         \tilde{\Psi'}(\widetilde{\theta}) = \sum_{j=1}^{m} \prod_{i=1}^{d}
         \widetilde{\theta}_{i}^{(\sigma_{i} - N t_{i}(j))}
        \end{equation}
         The solution is given by solving the following set of
         equations
         \begin{equation}
           \frac{\partial \widetilde{\Psi}'}{\partial \widetilde{\theta}_{i}} =0 \enspace,
           i=1,\ldots d .
         \end{equation}
         We have
         \begin{equation}
         \frac{\partial \widetilde{\Psi}'}{\partial \widetilde{\theta}_{i}} \in
         k[\widetilde{\theta}_{1}^{\pm},\ldots,
           \widetilde{\theta}_{d}^{\pm}] \enspace, i = 1, \ldots, d.
         \end{equation}
       \endproof

       In algebraic statistics, algebraic descriptions are used to
       analyze the maximum likelihood estimates of exponential
       models~\cite{PachterSturmfels:2005:AlgebraicStatisticsAndComputationalBiology}. In
       the view that maximum likelihood and maximum entropy are
       related,
       it will be interesting to compare these two methods from algebraic
       statistical point of view.


\section{Conclusion and Directions for Future research}
         \noindent
         In this paper we attempted to describe maximum (and hence minimum) entropy
         model in algebraic statistical framework. We showed that
         maximum entropy models are toric models when the constraint
         functions are assumed to be integer valued functions and the
         set of statistical models results from ME-principle is indeed
         an variety. In a dual estimation 
	 we
         demonstrated that when
         the information is in the form of empirical means, the
         calculation of ME-models can be transformed to solving set of
         Laurent polynomial equations. Work on computational algebraic
         algorithms for estimating ME-models are in progress. We hope
         that this will also shed light on possible interesting
         algebraic structures in information theoretic statistics.
         

        
\section*{References}

\bibliographystyle{jphysicsB}
\bibliography{papi}

\end{document}